\documentclass[sigplan,natbib=false]{acmart}

\AtBeginDocument{%
  }

\setcopyright{cc}
\setcctype{by}
\copyrightyear{2026}
\acmConference[Agent Skills '26]{AgentSkills '26 Workshop: ACM Conference on AI and Agentic Systems}{May 26, 2026}{San José, CA}
\acmBooktitle{Agent Skills '26 Workshop: ACM Conference on AI and Agentic Systems, May 26, 2026, San José, CA}

\acmDOI{}
\acmISBN{}

\usepackage[american]{babel}
\usepackage{csquotes}

\usepackage[T1]{fontenc}

\usepackage[most]{tcolorbox}
\usepackage{enumitem}
\usepackage{tikz}
\usepackage{graphicx}
\usepackage{tabularx}
\usepackage{booktabs}
\usepackage{multirow}
\usepackage{makecell}
\usepackage{pifont}

\newcommand{\roundframe}[1]{{\setlength\fboxrule{0pt}\fbox{\tcbox[colframe=black,colback=white,shrink tight,boxrule=0.5pt,extrude by=2.5pt]{\small #1}}}}
\tcbset{
  takeawaystyle/.style={
    colback=gray!5!white, %
    colframe=gray!75!black, %
    fonttitle=\bfseries\itshape, %
    title={#1}, %
    attach boxed title to top left={xshift=2mm, yshift=-3mm}, %
    boxed title style={
      colframe=white, %
      boxrule=0mm, %
      boxsep=1mm, %
    },
    enhanced, %
    boxrule=0.5mm, %
    top=3mm, %
  },
}
 \usepackage[table]{xcolor}
  \usepackage{xfp}        %
  \usepackage{siunitx} 
\newlength\MAX  
\newlength\SMMAX  

\usepackage{xspace}

\DeclareUrlCommand{\tturl}{\urlstyle{tt}}
\DeclareUrlCommand{\bftturl}{}
\DeclareUrlCommand{\ittturl}{}

\usepackage{doi}
\usepackage[sorting=anyt,%
            giveninits=true,%
            citestyle=numeric-comp,%
            style=acmnumeric,
            natbib=true,%
            backend=biber,%
            doi=true,%
            hyperref=true,%
            mincrossrefs=1000,
            date=year
            ]{biblatex}
\AtEveryBibitem{%
	\clearfield{editor}
	\clearfield{pages}
	\ifentrytype{book}{
	}{
		\ifentrytype{inbook}{}{%
			\clearlist{publisher}
			\clearname{editor}
			\clearfield{isbn}
			\clearfield{issn}
	}}
	\ifentrytype{inproceedings}{
		\clearfield{series}
		\clearfield{volume}
		\clearfield{edition}
		\clearfield{isbn}
		\clearfield{url}
		\clearfield{pages}
		\clearfield{languages}

	}{}
	\clearlist{location}
	\clearlist{address}
	\clearfield{language}
}

\renewbibmacro*{year}{%
  \iffieldundef{year}%
    {}%
    {\printfield{year}}%
}
\renewbibmacro*{date}{}

\usepackage{hyphenat}
\PassOptionsToPackage{hyphens}{url}
\usepackage{hyperref}

\mathchardef\UrlBreakPenalty=0
\mathchardef\UrlBigBreakPenalty=0

\hypersetup{
	colorlinks=true,
  linkcolor={red!40!black},
	citecolor={red!70!black},
	urlcolor={blue!80!black},
	breaklinks=true,
	filecolor=black,
	pdfpagelayout=SinglePage,
	unicode
}

\usepackage{cleveref}

\newif\ifreview

\begin{document}

\title{Context Matters: Repository-Aware Security Analysis of the Agent Skill Ecosystem}
\author{Florian Holzbauer}
\orcid{0000-0003-2494-0331}
\affiliation{%
  \institution{Interdisciplinary Transformation University (IT:U)}
  \city{Linz}
  \country{Austria}
}
\email{florian.holzbauer@it-u.at}

\author{David Schmidt}
\orcid{0009-0009-0119-1068}
\affiliation{%
  \institution{University of Vienna, CDL AsTra}
  \department{Faculty of Computer Science}
  \city{Vienna}
  \country{Austria}
}
\email{d.schmidt@univie.ac.at}

\author{Gabriel K. Gegenhuber}
\orcid{0000-0002-7225-6297}
\affiliation{%
  \institution{Interdisciplinary Transformation University (IT:U)}
  \city{Linz}
\country{Austria}
}
\email{gabriel.gegenhuber@it-u.at}

\author{Sebastian Schrittwieser}
\orcid{0000-0003-2115-2022}
\affiliation{%
  \institution{University of Vienna, CDL AsTra}
  \department{Faculty of Computer Science}
  \city{Vienna}
  \country{Austria}
}
\email{sebastian.schrittwieser@univie.ac.at}

\author{Johanna Ullrich}
\orcid{0000-0003-0297-9614}
\affiliation{%
  \institution{Interdisciplinary Transformation University (IT:U)}
  \city{Linz}
  \country{Austria}
}
\email{johanna.ullrich@it-u.at}

\renewcommand{\shortauthors}{Holzbauer et al.}

\begin{abstract}

Agent skills extend local AI agents, such as Claude Code and OpenClaw, with additional functionality. Their growing popularity has led to dedicated marketplaces resembling mobile app stores, as well as automated scanners that assess whether skills are benign or malicious. However, scanner reports from individual marketplaces classify up to 46.8\% of skills as malicious, raising concerns about false positives.
We present the largest empirical security analysis of the AI agent skill ecosystem to date. We collect 238,180 unique skills from three major distribution platforms and GitHub, and analyze their contents, behavior, and repository context. Unlike existing scanner-based assessments, which evaluate skills largely in isolation, our repository-aware analysis checks whether a flagged skill is consistent with its surrounding GitHub project.
This context substantially reduces the number of suspicious skills: only 0.52\% remain suspicious after repository-aware analysis. Our results show that existing scanners can substantially overestimate maliciousness when repository context is ignored. At the same time, we identify previously undocumented real-world attack vectors, including the hijacking of skills hosted in abandoned GitHub repositories. Overall, our findings provide a more robust view of the agent-skill ecosystem's current risk surface and highlight the need for context-aware security evaluation.

\end{abstract}



\maketitle
\section{Introduction}
Autonomous AI agents, such as Claude Code~\cite{claude_code} or OpenClaw~\cite{openclaw} extend large language models (LLMs) from standalone text generation systems into truly autonomous, closed-loop assistants that can plan, act, and learn complex tasks.
In a nutshell, the LLM interprets user requests, invokes adequate routines to serve them, and eventually integrates the results into subsequent reasoning steps.
A key concept are skills, which are reusable, modular components that extend an agent's capabilities like access to an external API, code execution, or data retrieval.
Following a standardized and open format~\cite{agentskills2025}, skills consist of natural language descriptions, informing the user and the LLM about its capabilities, in combination with executable logic implementing the functionality.
Skills are found on dedicated skill markets, e.g., ClawHub~\cite{steinberger:2026:clawhub}, Skill.sh~\cite{skills_sh}, and SkillDirectory~\cite{skilldirectory}, or traditional repositories like Github.
Agents might even discover them on their own, e.g., when reading on \url{moltbook.com}, a reddit-like platform for bots.

On the one hand, the tight integration of LLM reasoning with execution capabilities poses unique risks as anecdotal evidence of undesired email deletion emphasizes~\cite{chandonnet:2026:meta}.
A recurring pattern, on the other hand, are supply chain risks by integrating external resources for a system's extended functionality from market places or other sources.
Examples are machine images in compute clouds~\cite{bugiel:2011:amazonia}, docker hub for containers~\cite{shu:2017:study},  mobile applications for iPhones~\cite{schmidt:2026:supplychaininsecurityexposing}, or packet managers like npm and PyPI~\cite{zhang:2025:killing}, and nowadays, it appears, also skills for AI agents.
Among others, malicious skills attempted to steal private information from macOS~\cite{oliveira2026malicious} or redirect cryptocurrency assets~\cite{lakshmanan:2026:hackernews}.
In consequence, skill markets nowadays automatically scan the provided skills for security, and provide the results for orientation to their users. The total share of malicious skills, however, varies significantly among market places -- 46.8\% (ClawHub), 23.0\% (Skills.sh), and 6.0\% (SkillsDirectory).

In this paper, we present the largest empirical security study of the AI agent skill ecosystem, collecting and analyzing 238{,}180 skills from three distribution platforms and GitHub.
Our study revisits the high maliciousness rates reported by existing marketplaces and asks whether these classifications remain meaningful once skill and repository context are considered.
We structure the analysis around the following research questions.

\begin{itemize}[leftmargin=3em]

\item[\roundframe{RQ1}]
\textit{What skills are shared on marketplaces, and which new attack vectors emerge from the skill ecosystem?}

\item[\roundframe{RQ2}]
\textit{How do marketplaces and scanners classify skills as malicious?}

\item[\roundframe{RQ3}]
\textit{Can repository context improve existing security classifications of skills?}

\end{itemize}

We first characterize the collected skills across marketplaces and GitHub, including their scripts, embedded artifacts, and distribution structure.
We then analyze how marketplace scanners and the Cisco Skill Scanner~\cite{cisco_skill_scanner} classify skills as malicious and compare the consistency of their detections.
Finally, we reevaluate scanner-flagged skills by incorporating the surrounding GitHub repository context and measuring whether the repository documentation and code align with the skill specification.
This repository-aware analysis shows that isolated skill scanning substantially overestimates the ecosystem's risk.
At the same time, our broader analysis uncovers structural weaknesses in skill distribution platforms, including repository hijacking risks that allow adversaries to take over references to existing skills.
Summarizing, our paper contributes the following aspects:

\begin{itemize}

\item
\textbf{Large-scale ecosystem measurement.}
With 238{,}180 unique skills, we construct the largest cross-platform dataset of agent skills to date by collecting skills from three official marketplaces as well as GitHub repositories.
The dataset not only facilitates the analysis at hand, but also provides a basis for future longitudinal studies of the AI skill ecosystem.

\item
\textbf{Repository-aware skill analysis.}
Existing security scanners classify a large share of offered skills as malicious.
We conduct a semantic analysis that incorporates not only the skill description, but also the surrounding repository context.
Among 2{,}887 scanner-flagged skill and repository combinations, only 15 remain associated with suspicious repositories, corresponding to 0.52\%.
This substantially reduces the number of likely false positives and provides a more contextualized view of the ecosystem's risk surface.

\item
\textbf{Discovery of new attack vectors.}
We identify previously undocumented attack vectors in the AI skill ecosystem.
In particular, we demonstrate repository hijacking risks for seven abandoned repositories referenced by skill indexes, affecting 121 skills.
One affected skill has more than 1{,}000 recorded installations, a significant number considering the recency of the ecosystem.

\end{itemize}

To enable reproducibility and future work, we publish our code: \url{https://github.com/holzsec/repository-context-agentskills/}.

\section{Background and Related Work}

\paragraph{Agent and Skills.}
AI agents autonomously execute tasks in interaction with external services, and operate in a reasoning-action loop.
The LLM interprets a user request, selects and invokes an adequate capability, and eventually incorporates the results into subsequent reasoning.
Many agents support modular extensions, such as API access, code execution, or data retrieval, that are referred to as \emph{skills}. 
Skills typically come in a repository, combining a specification file (\texttt{SKILL.md}) with optional scripts, configuration files, or static assets. 
The specification file describes the skill's capabilities and its invocation context in natural language, enabling the autonomous agent to decide, while the latter represent the executable logic.
For interoperability, Anthropic recently specified a skill packaging format~\cite{skill_specification_mintlify}.

\paragraph{Skill Marketplaces.}
Agent skills are distributed over dedicated marketplaces such as ClawHub~\cite{steinberger:2026:clawhub}, Skills.sh~\cite{skills_sh}, and SkillsDirectory~\cite{skilldirectory}.
As of March 9, 2026, they provide 18,412, 86,800, and 36,109 skills, respectively.
Yet, the platforms differ, leading to different levels of control for the operators:
ClawHub curates and reviews uploaded skills, and hosts them itself.
Skills.sh adopts an open Git-based distribution model and indexes skills in external repositories.
SkillsDirectory also refers to external repositories, but moderates submissions and performs rule-based security scanning.

\paragraph{Security on Marketplaces.}
Malicious skills are known to manipulate agent behavior and after reports on their distribution over marketplaces~\cite{oliveira2026malicious,lakshmanan:2026:hackernews,kellner:2026:synk,schmotz2026skill}, marketplaces nowadays automatically scan the offered skills for security.
The results, typically a classification of whether a skill is benign or malicious in combination with a short explanation on the reasoning, are provided as metadata to their users.
Therefore, ClawHub relies on VirusTotal~\cite{virustotal} and a custom LLM-based detection system, Skills.sh integrates several third-party scanners, and SkillsDirectory reports the use of more than 50 rule-based detection mechanisms. 
The share of skills reported as suspicious varies substantially across marketplaces, namely 46.8\% (ClawHub), 23\% (Skills.sh), and 6\% (SkillsDirectory). 
Across all marketplaces, the share remains high, which indicates either a large number of malicious skills or a high false positive rate.

\paragraph{Empirical Studies on the Skill Ecosystem.}
Studies by third parties come to a similar share of malicious skills.
An analysis of 3,984 skills on ClawHub and skills.sh~\cite{kellner:2026:synk} found 13.4\% of them having a critical-level security flaw like malware distribution, or prompt injection, and 36.82\% show (more minor) security pitfalls like hard-coded API keys or insecure credential handling.
Another analysis investigated 31,132 skills~\cite{liu:2026:agent} from skills.rest and skillsmp.com, and found that 26.1\% of the skills contain a security vulnerability such as prompt injection, and data exfiltration. %
The largest study by now investigated 40,285 skills from skills.sh~\cite{ling:2026:agent}.
While predominantly focusing on their publication behavior over time as well as prompt length, the author also assessed their security and concluded 9\% of them having critical flaws.
In this context, also multiple skill scanners emerged, e.g., SkillScan~\cite{liu:2026:agent},  SkillFortify~\cite{bhardwaj:2026:formal}, Snyk~\cite{kellner:2026:snyk_report}, and the Cisco skill scanner~\cite{cisco_skill_scanner}.
Also, multiple works promise annotated data sets of skills for security benchmarking~\cite{bhardwaj:2026:formal,liu:2026:agent}.
However, upon inspection, we were unable to find them, impeding a direct comparison of our analysis with those from previous work on the same data sets.

\paragraph{Security of AI Agents.}
Instead of skills, vulnerabilities might also directly affect the agent.
(Meanwhile fixed) \textit{ClawJacked} enabled an attacker to gain control over Open Claw instances using a web socket to localhost~\cite{lakshmanan:2026:hackernews}.
Alternatively, attackers could use web sockets to modify log files that the AI agents eventually rely on for troubleshooting~\cite{steinberger:2026:openclaw_github_advisory}.
Via prompt injection, OpenClaw was persuaded to reveal private keys~\cite{cruz:2026:openclaw_keys}.
In the manner of social engineering, moltbook appears to be exploited to extract metadata for reconnaissance.
The programming agent Claude Code was meanwhile also vulnerable to remote code execution~\cite{CVE202559536}, and revealed API keys~\cite{CVE202621852}.
Finally, Shodan currently discovers 55,561 of such OpenClaw instances on the Internet~\cite{shodan:2026:openclaw_instances}, and a honeypot provides a glimpse into the attackers' current strategies~\cite{fogel:2026:caught_in_the_wild}.

\paragraph{Scientific Literature.}
Due to the recency of the topic, most reports appear in non-scientific venues, e.g., blogs or as unaccepted preprints.
Yet, scientific literature already discusses autonomous AI agents~\cite{acharya:2025:agenticai,pati:2025:agenticai,basu:2026:openclaw}.
Researchers consider security both a potential application of these systems and a major challenge for them.
On the one hand, autonomous AI can continuously monitor malicious activities and immediately block them, which can improve security.
On the other hand, these systems can themselves pose security threats because they combine autonomy with access to large and potentially sensitive datasets.
A dedicated survey on security~\cite{deng:2025:artificial} classified the threat landscape of AI agents, also including supply chain threats.

\section{Methodology}

\begin{figure}[t]
    \centering
    \includegraphics[width=\linewidth]{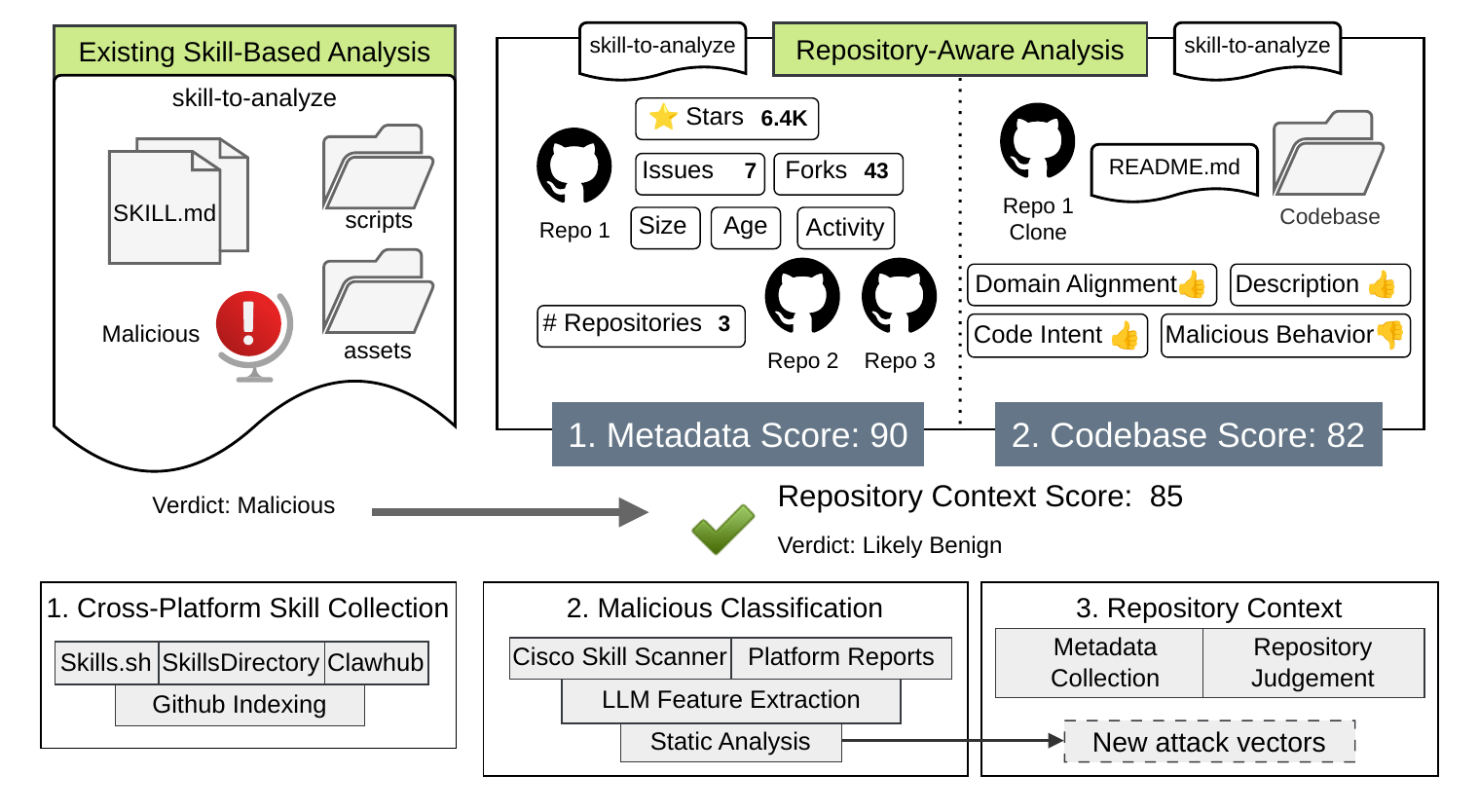}
    \caption{Overview of our repository-aware skill analysis to reduce the high number of false positives. Our approach consists of three stages, encompassing skill collection, malicious classification, and repository context analysis.}
    \Description{The figure illustrates our repository-aware skill analysis pipeline designed to reduce malicious claims. It consists of three sequential stages. First, the system collects skills from multiple platforms in a cross-platform aggregation step. Second, it applies a classification component to identify potentially malicious skills based on predefined criteria. Third, it performs repository context analysis by examining associated code repositories to validate claims and detect inconsistencies.}
    \label{fig:methodology}
\end{figure}

We study the security of agent skills using a three-stage measurement pipeline, illustrated in \Cref{fig:methodology}. 
Our methodology is designed to capture the breadth of the emerging skill ecosystem, compare how existing tools assess skill risk, and evaluate whether repository context can help interpret scanner alerts. 
Rather than treating any individual scanner as ground truth, we use scanner outputs as security signals that require contextual interpretation.

\subsection{Cross-Platform Skill Collection}

To answer \roundframe{RQ1}, we collect agent skills from multiple sources, including public skill marketplaces and GitHub repositories. 
This cross-platform collection allows us to study both curated or indexed marketplace skills and skills that are published independently by developers. 
Since marketplaces differ in how they host and reference skills, we normalize all collected artifacts into a common representation based on the skill directory and its associated files.

To extend coverage beyond known marketplaces, we search public GitHub activity data for repositories likely to contain skill definitions. 
Candidate repositories are cloned and scanned for \texttt{SKILL.md} files, which serve as the entry point for identifying skill artifacts. 
We apply resource limits during collection to ensure scalability and deduplicate skills using content hashes to avoid counting identical artifacts multiple times.

After collection, we perform static analysis over each skill artifact. 
We enumerate contained files, record file types and directory structure, and extract ecosystem-level properties such as the presence of executable scripts. 
We further scan skill artifacts for embedded secrets and validate detected credentials where possible. 
This analysis provides the basis for characterizing the structure, content, and potential exposure risks of the skill ecosystem.

\subsection{Scanner-Based Skill Assessment}

To answer \roundframe{RQ2}, we evaluate how existing security mechanisms classify agent skills. 
We collect scanner reports exposed by skill marketplaces and complement them with platform-independent analyses. 
This includes an open-source skill scanner and an LLM-based behavioral feature extraction pipeline.

The scanner-based analysis serves two purposes. 
First, it allows us to measure how frequently skills are flagged by different tools. 
Second, it enables us to compare the consistency of these tools by analyzing overlap between their detections. 
Because the ecosystem lacks a reliable ground-truth dataset of malicious skills, we interpret scanner results as alerts rather than definitive labels. 
This distinction is important because different scanners rely on different assumptions, rules, and behavioral models, which can lead to inconsistent classifications.

Our LLM-based analysis extracts structured behavioral indicators from each skill.  Because the full skill contents exceeded the LLM's context window, the analysis used a LLM-based auxiliary script to extract relevant information from the skill files. 
The prompt evaluates whether a skill exhibits properties associated with risky behavior, such as system interaction, network communication, credential handling, persistence, or abuse potential. 
These features complement deterministic scanner outputs and provide a platform-independent basis for comparing skills across marketplaces and repositories.

\subsection{Repository-Aware Contextual Analysis}

We analyze whether skills flagged by scanner-based methods remain suspicious when evaluated in the context of their surrounding repositories to answer \roundframe{RQ3}. 
This step is motivated by the fact that scanners typically inspect skills in isolation, although many skills are embedded in larger projects whose documentation, source code, and development history can explain their behavior.

We therefore perform repository-aware analysis for high-risk scanner alerts where GitHub repository context is available. 
For each selected skill, we collect repository metadata and relevant codebase context. 
We then derive two complementary signals. 
The first captures codebase alignment, i.e., whether the skill's described functionality is consistent with the surrounding repository documentation and implementation. 
The second captures repository maturity, based on metadata such as project age, activity, size, and popularity.

We combine these signals into a repository-context score. 
The score is intended as a contextual trust and triage signal rather than a definitive maliciousness label. 
A low score indicates that a flagged skill is weakly supported by its repository context, while a high score indicates that the repository provides evidence that the flagged behavior may be expected or benign in context. 
For skills appearing in multiple repositories, we aggregate context scores across representative repositories to account for the fact that the same skill may appear in different environments.

Finally, we validate the repository-aware analysis through manual inspection of a sample of flagged repositories. 
This validation assesses whether repositories identified by scanners as suspicious appear benign or suspicious to independent reviewers and whether the automatically derived context scores are consistent with human judgments.

\section{RQ1: Cross-platform Skill Analysis}

We first present results on the differences between the marketplaces and the additional skills that we discovered on GitHub. We then provide an overview of the content of the published skills, and demonstrate how attackers can hijack skills referenced in two marketplaces.

\begin{table}[t]
\setlength{\tabcolsep}{3pt}
\caption{Overview of collected agent skills from ClawHub~\cite{steinberger:2026:clawhub}, SkillDirectory~\cite{skilldirectory}, Skills.sh~\cite{skills_sh}, and GitHub. Retrieved denotes successfully downloaded skills retained for analysis; Added denotes skills retained after cross-source deduplication. For Skills.sh, the crawl retrieved 55{,}366 listed skills and 77{,}456 additional skills extracted from referenced repositories.}
\label{tab:clawhub_overview}
\centering
\begin{tabularx}{\linewidth}{Xrrrr}
\toprule
\textbf{Skill Metric} & \textbf{ClawHub} & \textbf{SkillsDir.} & \textbf{Skills.sh} & \textbf{GitHub} \\
\midrule
\textbf{Indexed}
& 16{,}755
& 32{,}896
& 79{,}735
& 142{,}824
\\
\textbf{Retrieved}
& 16{,}755
& 17{,}611
& 125{,}928
& 136{,}095
\\
\textbf{Added}
& 16{,}755
& 17{,}611
& 112{,}231
& 91{,}583
\\\addlinespace
\textbf{Owners}
& n/a
& 709
& 7{,}950
& 14{,}197
\\
\textbf{Repositories}
& n/a
& 766
& 9{,}431
& 16{,}413
\\
\midrule
\multicolumn{5}{l}{\textbf{\#Total} 238{,}180 (distinct analyzed skills)} \\
\bottomrule
\end{tabularx}
\end{table}

\begin{figure}[t]
\centering
\begin{minipage}{0.43\linewidth}
    \centering
    \includegraphics[width=1\linewidth]{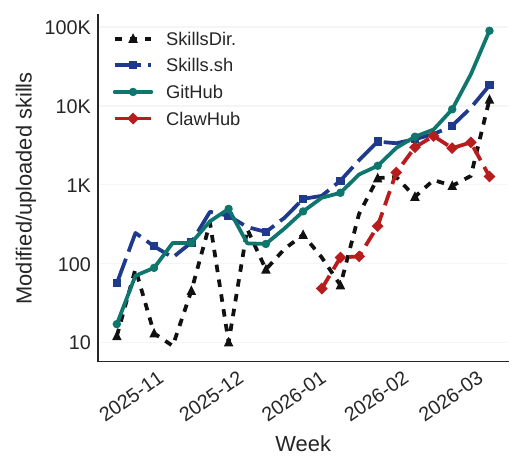}
    \caption{All platforms show an increase in number of weekly updated agent skills over time.}
    \Description{The figure shows the temporal evolution of weekly modified agent skills across all investigated platforms. Each platform exhibits a clear upward trend, indicating a steady increase in the number of newly introduced skills over time.} 
        \label{fig:weeklyskills}
\end{minipage}
\hfill
\begin{minipage}{0.53\linewidth}
    \centering
    \includegraphics[width=1\linewidth]{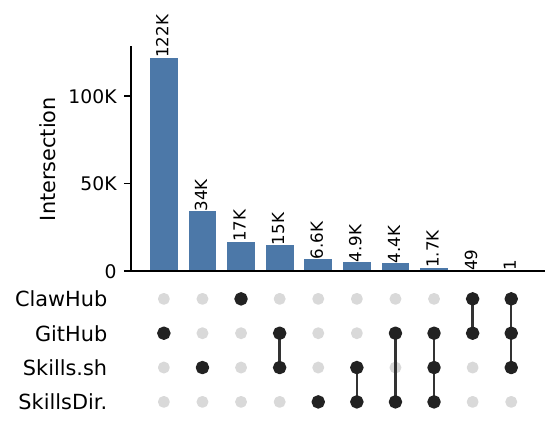}
    \caption{Overlap of skills retrieved from the marketplaces.}
    \Description{The figure illustrates the overlap of skills published across different marketplaces.}
    \label{fig:overlap}
\end{minipage}
\end{figure}

\subsection{Collected Skill Dataset}

\Cref{tab:clawhub_overview} summarizes the skills that we crawled from the different marketplaces and the subset that we successfully downloaded and analyzed.
In total, we indexed 16{,}755 skills from ClawHub, 79{,}735 marketplace-listed skills from Skills.sh, 32{,}896 skills from SkillDirectory, and 142{,}822 skills referenced on GitHub.
For Skills.sh, 55{,}366 of the indexed marketplace entries could be retrieved directly.
In addition, we extracted 77{,}456 further skill folders from repositories referenced by Skills.sh, resulting in 125{,}928 analyzed Skills.sh skills in total.

During collection, we encountered several issues that limited the number of retrievable skills.
For SkillDirectory, a large repository subdirectory that effectively represents a separate skill collection was omitted, resulting in 13{,}304 skills not being retrieved.
In addition, changes in repository structures caused skill paths referenced by the marketplaces to become invalid.
Specifically, 21{,}800 skills referenced by marketplace indexes were no longer present in the repositories, and another 3{,}991 repositories no longer contained skills referenced by \textit{Skills.sh}.
Furthermore, 188 repositories had become private or required authentication at the time of collection.
These issues highlight the dynamic nature of the ecosystem and the challenges of reliably archiving skill datasets.
\Cref{fig:weeklyskills} further shows the number of skills added to these platforms each week since November 2025.

In total, we successfully downloaded and analyzed 16{,}755 skills from ClawHub, 125{,}928 from Skills.sh, 17{,}611 from SkillDirectory, and 136{,}095 from GitHub.
In \Cref{fig:overlap}, we show the overlap of skills across the platforms.
The figure illustrates that a substantial fraction of skills appears on multiple marketplaces, indicating that many platforms reference the same underlying repositories.
In particular, GitHub acts as the primary hosting platform for most skills, while the marketplaces serve as discovery layers.
At the same time, each marketplace contributes skills not listed in the others, reflecting differences in indexing scope and update frequency.
Overall, aggregating multiple marketplaces increases coverage of the skill ecosystem and results in a final dataset of 238{,}180 cross-marketplace unique skills for further analysis.

The marketplaces also differ in their ecosystem structure.
Skills.sh references 7{,}950 owners and 9{,}431 repositories, while the GitHub dataset spans 14{,}197 owners and 16{,}413 repositories.
SkillDirectory is comparatively smaller, with 709 owners and 766 repositories.
These numbers indicate that many repositories host multiple skills, suggesting that developers frequently group related skills within a single project rather than publishing them individually.

\paragraph{Skill Content.}
We further analyzed which scripts skills contain and provided a table in our artifact~\cite{artifact:skill_scripts}. 
Across all marketplaces, Python scripts appear most frequently, followed by shell scripts, JavaScript, and TypeScript. However, the share of skills that include at least one script differs across marketplaces. While Skills.sh, SkillsDirectory, and GitHub have a similar range of skills containing scripts (11.8\% to 15.7\%), ClawHub shows a substantially higher share of skills including at least one script (44.1\%). One explanation for this difference could be that ClawHub more specifically targets OpenClaw instead of general agents.

We further evaluate whether scripts reside in a directory named \texttt{scripts/}, as defined by the specification~\cite{skill_specification_mintlify}. Again, ClawHub represents an outlier. Among skills that include scripts from ClawHub, 13.2\% lack a \texttt{scripts/} directory. In contrast, only 2.9\% to 3.4\% of skills from the other marketplaces contain scripts without the corresponding directory.

\paragraph{Secrets.}
We further analyzed whether skills contain valid tokens and credentials. In total, we discovered 12 functional credentials, including four for the NVIDIA API, three for ElevenLabs, two Gemini tokens, two MongoDB credentials, and one credential each for Amplify, Postgres, and X AI.
Attackers could abuse these credentials to access third party services and perform actions on behalf of the credential owner. For example, NVIDIA, ElevenLabs, Gemini, and xAI token allow access to AI services, which attackers could use to issue requests that incur costs for the owner. One reason for the relatively small number of discovered secrets may be that most skills are hosted on GitHub. In contrast to mobile apps~\cite{schmidt:2025:app_secrets} or accessible storage buckets~\cite{yadmani:2025:s3_secrets}, developers are likely more aware that the code is publicly visible and that attackers could access any embedded secret.

\subsection{Skill Provisioning}
For the security of distributed skills, similar considerations as for dependency management systems apply. Attackers can hijack dependencies if the system does not host the dependency itself and the URL hosting it can be taken over, for example because a username on GitHub was renamed~\cite{schmidt:2025:app_secrets}. In addition, the authentication mechanism used to publish a skill plays an important role, as weak or missing authentication can enable attackers to hijack existing dependencies or skills. We therefore looked into the currently implemented authentication mechanisms for publishing skills and their distribution. For authentication, ClawHub and SkillsDirectory rely on GitHub authentication, while Skills.sh provides no authentication. Instead, the marketplace adds skills when users download them using the command line tool with telemetry enabled. In this sense, the system resembles Go modules, which also do not implement a separate authentication mechanism but cache dependencies~\cite{gu:2023:software_registries}.
However, instead of caching or redistributing the skills, Skills.sh directly downloads them from GitHub. This design can enable attackers to hijack existing repositories if the previous owner renames their account and the repository has not yet reached the required download threshold that would cause GitHub to retire the repository name~\cite{schmidt:2026:supplychaininsecurityexposing}. The same issue also affects SkillsDirectory. Although SkillsDirectory provides the option to download skills from its website, the command line tool currently attempts to download the skill from GitHub.
In contrast, ClawHub directly distributes the skill. This design reduces the dependency on third-party URL management and therefore decreases the risk of repository hijacking.

\paragraph{Skills Vulnerable to Hijacking.}
To test whether GitHub mitigates repository hijacking for vulnerable skills, we created test accounts using the associated usernames and entered the repository names without creating the repositories. This approach keeps existing redirects functional while revealing whether an attacker could recreate the repository under the same name~\cite{schmidt:2026:supplychaininsecurityexposing}. To prevent attackers from hijacking vulnerable skills, we keep the account names associated with vulnerable skills reserved. We performed this step for all identified repositories with five or more stars, as the process of registering GitHub accounts cannot be automated.

Overall, we discovered 121 skills that forward to seven vulnerable repositories. Among them, 77 skills indexed by Skills.sh reference five vulnerable repositories, while 44 skills listed on SkillsDirectory reference two additional vulnerable repositories. One hijackable repository referenced by SkillsDirectory has 159 stars, whereas the maximum star count among vulnerable repositories referenced by Skills.sh is 48. 
Using the download statistics provided by Skills.sh, we further assessed how frequently hijackable skills were downloaded. The median number of downloads is 25, while the most often downloaded skill reached 2,032 downloads. 

We responsibly disclosed this attack vector to the affected platforms and recommended switching to a direct distribution model similar to OpenClaw.

\paragraph{New Ecosystems, Old Issues.}
Based on the source code of ClawHub~\cite{clawhub_code}, we implemented a crawler for skill and security reports. During this process, we discovered that the associated endpoint returns additional owner metadata. In particular, the API exposes the email address associated with each user's GitHub account. This information is not shown by default on GitHub profiles and is also not visible through the ClawHub website. Therefore, we did not expect the ClawHub API to disclose this data.

\section{RQ2: Malicious Classification}

To answer \roundframe{RQ2}, we study how existing security scanners classify agent skills and evaluate the consistency of their maliciousness assessments. We compare scanner reports from skill marketplaces with the results of our own analysis pipeline, which includes the Cisco Skill Scanner and an LLM-based behavioral classifier. This comparison allows us to quantify how different scanning approaches interpret the same skills and to identify potential overclassification of malicious behavior. Understanding these differences is important because high false positive rates may reduce trust in the ecosystem, confuse end-users, and motivate the need for repository-aware analysis.

\paragraph{Malicious Classification Rates.}

\begin{table}[t]
    \centering
    \caption{Comparison of security scanners used in the skill ecosystem. The table contrasts scanners deployed on ClawHub and Skills.sh (highlight in gray) with Cisco's skill scanner and our LLM-based feature set. } %
    \label{tab:skills_sh_scanner_comparison}

    \setlength{\tabcolsep}{2pt}
    \begin{tabularx}{\linewidth}{lXrrrr}
        \toprule
        \textbf{} & \textbf{Scanner} & \textbf{Scanned} & \textbf{Pass} & \textbf{Fail} & \textbf{Fail Rate} \\
        \midrule
        \multirow[c]{5}{*}{\rotatebox{90}{Clawhub}} &
        \cellcolor{gray!20} VirusTotal            & \cellcolor{gray!20} 12,213           & \cellcolor{gray!20}7,792                   & \cellcolor{gray!20}4,421         & \cellcolor{gray!20}36.20\%                            \\
        & \cellcolor{gray!20} OpenClaw Scanner      & \cellcolor{gray!20} 14,244           & \cellcolor{gray!20}8,271                   & \cellcolor{gray!20}5,973         & \cellcolor{gray!20}41.93\%                            \\
        \addlinespace
        & GPT 5.3-based        &   16,424        &   10,050  & 6,374 & 38.8\%    \\
        & Cisco Skill Scan  &   16,745  &  13,941    & 2,804 & 16.74\%     \\ \midrule

        \multirow[c]{6}{*}{\rotatebox{90}{Skills.sh}} &
         \cellcolor{gray!20}agent-trust-hub       & \cellcolor{gray!20}62,163           & \cellcolor{gray!20}53,611                  & \cellcolor{gray!20}8,552         & \cellcolor{gray!20}13.76\%                            \\
        & \cellcolor{gray!20}snyk                  & \cellcolor{gray!20}46,414           & \cellcolor{gray!20}42,843                  & \cellcolor{gray!20}3,571         & \cellcolor{gray!20}7.69\%                             \\
        & \cellcolor{gray!20}socket                & \cellcolor{gray!20}56,695           & \cellcolor{gray!20}54,544                  & \cellcolor{gray!20}2,151         & \cellcolor{gray!20}3.79\%                             \\
        \addlinespace
        & GPT 5.3-based         & 52,577           & 38,234                  & 14,343        & 27.28\%                            \\
        & Cisco Skill Scan      & 52,577           & 45,196                  & 7,381         & 14.04\%                            \\
        \bottomrule
    \end{tabularx}
\end{table}

We compare the malicious classification rates reported by existing marketplace scanners with the results of our own analysis pipeline. Across both platforms, we observe substantial differences between scanners. On Clawhub, the OpenClaw scanner flags up to 41.93\% of skills as suspicious, while VirusTotal reports a similar rate of 36.20\%. Our GPT-5.3 based approach produces comparable results, classifying 38.8\% of skills as malicious. In contrast, the Cisco Skill Scanner reports a significantly lower fail rate of 16.74\%. These differences highlight that the perceived security of the skill ecosystem strongly depends on the chosen scanning approach.

The discrepancy between scanners is more pronounced when comparing the two marketplaces. On Clawhub, fail rates range from 16.7\% to 41.9\%, suggesting that a substantial fraction of skills may exhibit potentially suspicious behavior. 
In contrast, scanners deployed on Skills.sh report much lower fail rates, ranging from 3.79\% to 13.76\%. Our own analysis tools show similar trends: the GPT-5.3-based analysis classifies 27.28\%  of Skills.sh skills as suspicious, while the Cisco scanner flags 14.04\%. These discrepancies indicate that existing scanners produce inconsistent classifications when they analyze skills in isolation. In particular, scanners that rely on behavioral heuristics or language model reasoning flag substantially larger portions of the ecosystem as suspicious. Such elevated fail rates can reduce trust in the ecosystem and suggest that many skills are misclassified as malicious. This observation motivates the need for additional context to enable more accurate classification of skill behavior.

\begin{figure}[t]
  \centering
  \begin{minipage}{0.55\linewidth}
      \centering
      \includegraphics[width=\linewidth]{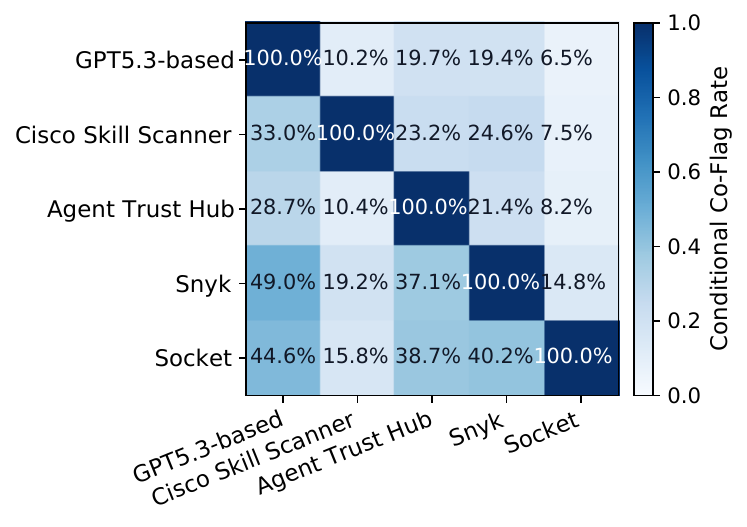}
      \caption{Conditional scanner agreement on Skills.sh common skills.}%
  \Description{The figure presents the conditional agreement between scanners on common skills from Skills.sh.}
      \label{fig:skillssh_conditional_overlap}
  \end{minipage}
  \hfill
  \begin{minipage}{0.40\linewidth}
      \centering
      \includegraphics[width=\linewidth]{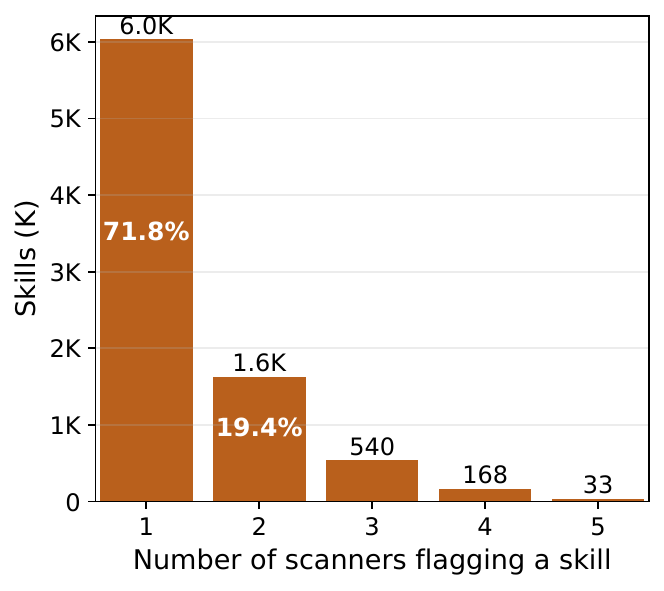}
      \caption{Number of Skills.sh common skills flagged by exactly \(k\in\{1,
  \dots,5\}\) scanners.}
        \Description{The figure shows the number of Skills.sh common skills flagged by exactly \(k\in\{1,
  \dots,5\}\) scanners.} %
      \label{fig:skillssh_flagged_by_k}
  \end{minipage}
  \end{figure}

\paragraph{Cross-Scanner Agreement.} To evaluate how consistently scanners classify skills as malicious, we compare the results of five scanners on the subset of 27,111 Skills.sh skills analyzed by all tools. \Cref{fig:skillssh_conditional_overlap} shows the conditional overlap between scanners, expressed as the probability that a skill flagged by scanner \(A\) is also flagged by scanner \(B\). Overall, agreement between scanners is low and often asymmetric. For example, 33\% of skills flagged by the Cisco Skill Scanner are also flagged by the GPT-5.3-based analysis, whereas only 10.2\% of GPT-5.3 detections overlap with Cisco. Similar patterns appear across other scanner pairs, indicating that scanners frequently identify different sets of skills as suspicious. \Cref{fig:skillssh_flagged_by_k} illustrates the distribution of detections across scanners.
Among the 8,402 skills flagged by at least one scanner, most are flagged by a single scanner, 6,032 skills. In contrast, 1,629 skills are flagged by two scanners and 540 by three scanners. Only 168 skills are flagged by four scanners, and just 33 skills are flagged by all five scanners. The limited overlap shows that scanner consensus is rare and that most detections lack corroboration by other tools.

\paragraph{Repository-level Classification Rates.}

\begin{table}[t]
\centering
\caption{Skill.sh malicious detection rates for skills and repositories (skill flagged if at least one scanner flagged, repository malicious if at least one skill is malicious)}
\setlength{\tabcolsep}{3pt}
\label{tab:skills_sh_detection_rates_strict}
\begin{tabularx}{\linewidth}{Xrrr}
\toprule
\textbf{Category} & \textbf{Total} & \textbf{Flagged} & \textbf{Rate} \\
\midrule
Skills (overall) & 62,219 & 12,004 & 19.29\% \\
Skills (repo stars > 1000) & 7,725 & 1,656 & 21.44\% \\
Skills (installs > 1000) & 755 & 122 & 16.16\% \\ \midrule
Repositories (overall) & 8,451 & 3,878 & 45.89\% \\
Repositories (stars > 1000) & 528 & 268 & 50.76\% \\
Repositories (installs > 1000) & 171 & 105 & 61.40\% \\
\bottomrule
\end{tabularx}
\end{table}

To better understand how scanner results translate from individual skills to repositories, we aggregate skill-level detections at the repository level. \Cref{tab:skills_sh_detection_rates_strict} shows the resulting malicious classification rates for both skills and repositories on Skills.sh. A skill is considered malicious if at least one scanner flags it, while a repository is classified as malicious if any of its contained skills is flagged.
Overall, 19.29\% of skills are flagged as malicious. When focusing on popular skills, the rates remain comparable: 21.44\% for skills hosted in repositories with more than 1,000 stars and 16.16\% for skills with more than 1,000 installs. However, the picture changes when aggregating these detections at the repository level. Nearly half of all repositories (45.89\%) contain at least one flagged skill. This proportion increases further for popular repositories, reaching 50.76\% for repositories with more than 1,000 stars and 61.40\% for repositories associated with highly installed skills.
These results suggest that even well-known repositories are frequently classified as malicious when applying strict aggregation rules. The effect is driven by repositories containing multiple skills: as the number of skills per repository increases, the probability that at least one skill is flagged also increases. Consequently, repository-level aggregation can substantially amplify skill-level detections and may overstate the prevalence of malicious repositories in the ecosystem.

\section{RQ3: Repository-Aware Analysis}

To answer \roundframe{RQ3}, we analyze whether scanner flagged skills remain suspicious when evaluated in the context of their surrounding GitHub repositories. We focus on skills flagged by both the Cisco Skill Scanner, with severity high or critical, and our GPT-5.3 based analyzer, with a score greater than 3. This yields 8,153 flagged skill and repository combinations.

We exclude ClawHub skills because they lack GitHub repository context, and skills located in repository roots because they only allow metadata based analysis. From the remaining set, we randomly sample 3,000 skill and repository combinations and collect repository metadata together with full repository clones. Cloning failed in 113 cases, leaving 2,887 combinations for codebase evaluation.

\paragraph{Metadata Score.}

\begin{figure}[t]
  \centering
  \begin{minipage}{0.95\linewidth}
      \centering
    \includegraphics[width=0.89\linewidth]{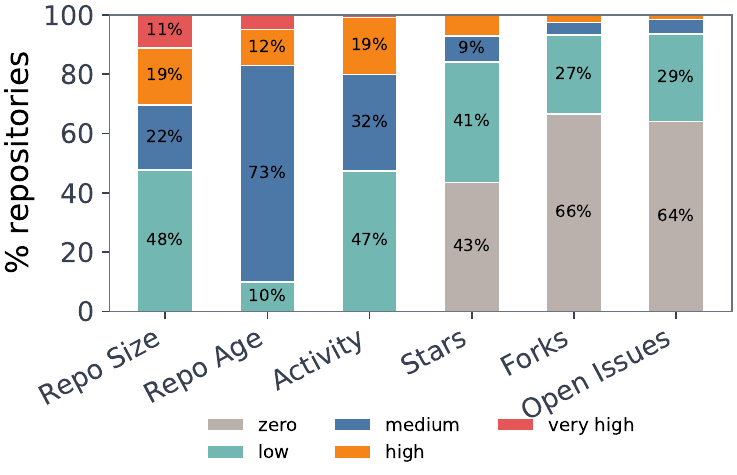}
    \caption{Metadata scores for unique repositories.}
    \Description{The figure presents the metadata scores for unique repositories.}
    \label{fig:metadata_buckets}
  \end{minipage}
\end{figure}

We first analyze repository metadata, including size, age, activity, popularity, and issue activity. \Cref{fig:metadata_buckets} shows that repositories containing flagged skills are typically small and have limited popularity: 47.6\% are smaller than 2\,MB, 43.4\% have no stars, 66.5\% have no forks, and 64.0\% have no open issues. At the same time, many repositories remain active, with 47.4\% updated within the last week. Compared to a random set of 1,500 repositories with matching marketplace distribution, repositories containing flagged skills do not exhibit distinct metadata characteristics. This suggests that metadata differences are driven more by marketplace composition than by suspicious skills.

\paragraph{Codebase Score.}

To evaluate whether a flagged skill aligns with its repository, we use an evidence based prompt that considers domain alignment, code similarity, README consistency, support signals, and repository level maliciousness. To control analysis cost, we limit the repository context to up to 200 lines from the \texttt{SKILL.md} and README files and up to three repository files with 100 lines each.

Most repositories provide sufficient context: 94.1\% contain a README, 65.7\% contain code, and 61.9\% contain both. Repository context often supports the skill purpose: domain matching is high for 45.9\% of skills and medium for 26.1\%, meaning that roughly 72\% show at least moderate thematic alignment. Direct code alignment is lower, with 9.9\% high and 28.6\% medium similarity. Repository level maliciousness is rare, with 98.0\% of repositories in the lowest maliciousness category and only two repositories showing high maliciousness signals. These results suggest that many scanner alerts are false positives caused by analyzing skills without repository context.

\begin{figure}[t]
    \centering
    \includegraphics[width=0.95\linewidth]{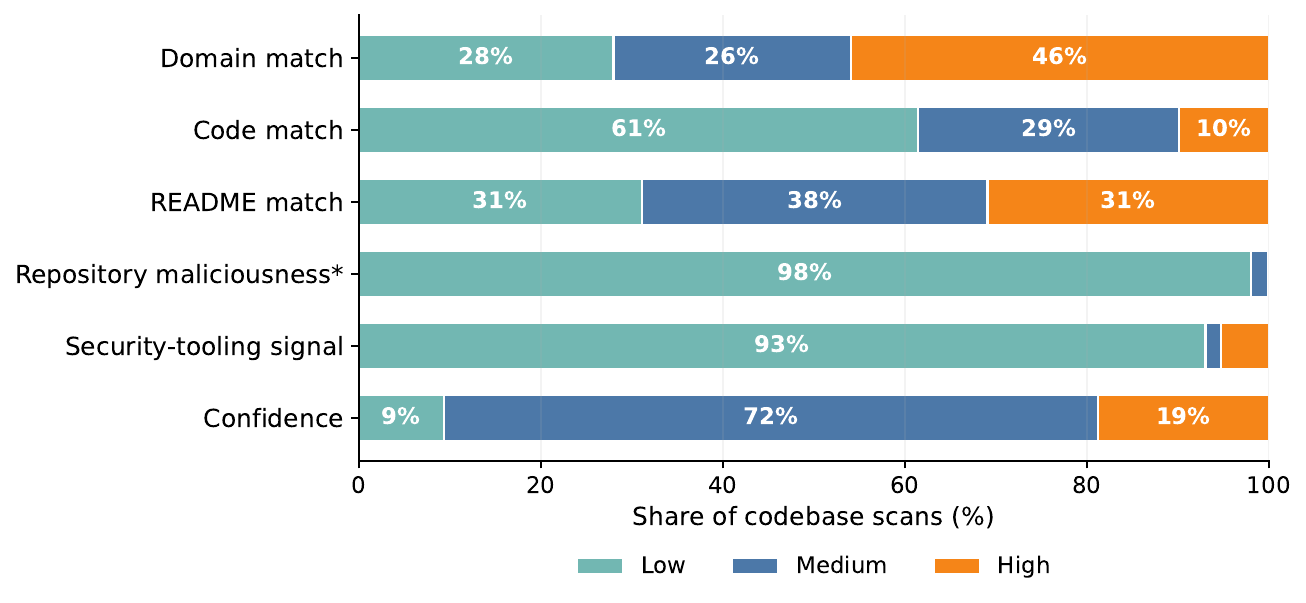}
    \caption{Codebase score category results. Categories marked with * indicate that lower values are better.}
    \Description{The figure shows the codebase score results per category.}
    \label{fig:codebase_categories}
\end{figure}

\paragraph{Repository Context Score.}

\begin{figure}[t]
    \centering
    \includegraphics[width=\linewidth]{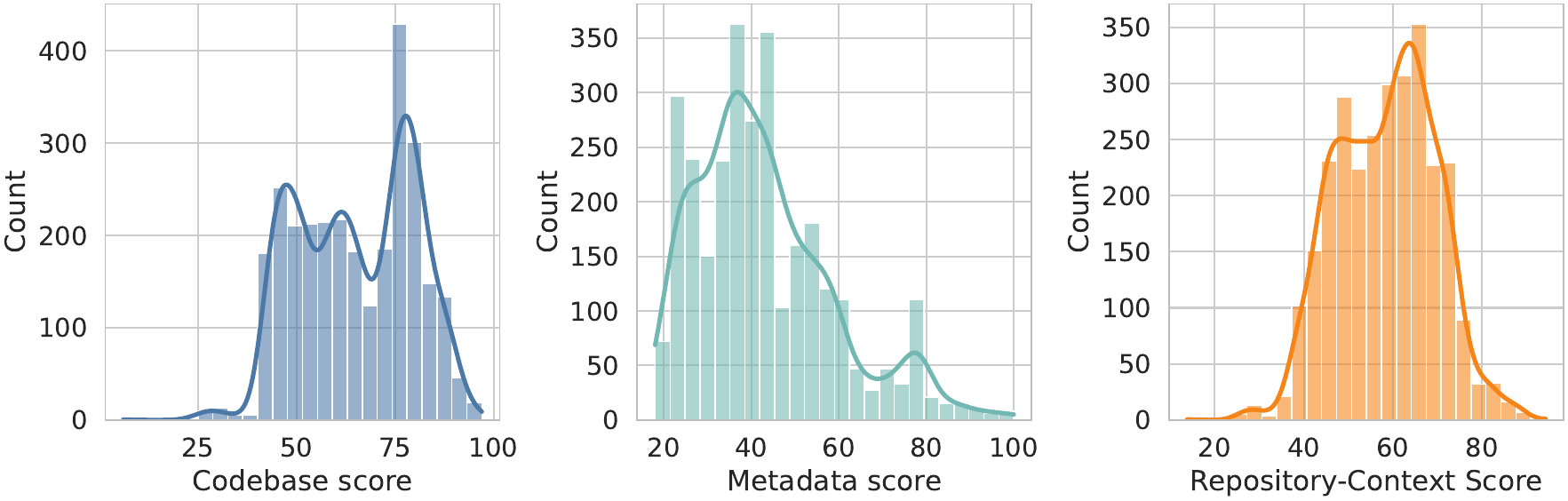}
    \caption{Repository context score and its weighted subcomponents, with 70\% codebase score and 30\% metadata score.}
    \Description{The figure illustrates the repository context score together with its weighted subcomponents.}
    \label{fig:score_panel}
\end{figure}

We combine the codebase score and metadata score into a repository context score, weighted 70\% and 30\%, respectively. As shown in \Cref{fig:score_panel}, the combined score has a mean of 58.5. The codebase score is higher, with a mean of 65.1, while the metadata score is lower, with a mean of 42.9, reflecting the young and low popularity repositories in the ecosystem.

We divide the repository context score into three categories. Only 121 cases (4.2\%) fall below 40, indicating weak repository embedding or low repository maturity. The largest groups are the intermediate category from 40 to below 60, with 1,373 cases (47.6\%), and the high category above or equal to 60, with 1,393 cases (48.3\%). Overall, most scanner flagged skills show moderate to strong repository linkage.

\paragraph{Suspicious Repositories.}

After repository aware analysis, only 15 skill and repository combinations remain suspicious. These cases correspond to repositories where the skill aligns with the codebase, but the repository itself appears suspicious and is not categorized as a security tool. This represents 0.52\% of the 2,887 evaluated combinations. These repositories have similar metadata scores to the full set, 41.1 versus 42.9, but lower codebase scores, 51.5 versus 65.2. This shows that repository context substantially reduces the suspicious set while still separating suspicious environments from benign scanner alerts.

\paragraph{Validation.}

Because no public ground truth dataset labels repositories as malicious with respect to agent skills, we validate our results through manual inspection. Two independent researchers reviewed 20 randomly sampled repositories from the flagged set. Of the 18 repositories that were still available, both reviewers classified all as benign based on documentation, code structure, and functionality. This supports our interpretation that many scanner flagged skills do not appear malicious once their repository context is considered.

\begin{figure}[t]
    \centering
    \includegraphics[width=0.8\linewidth]{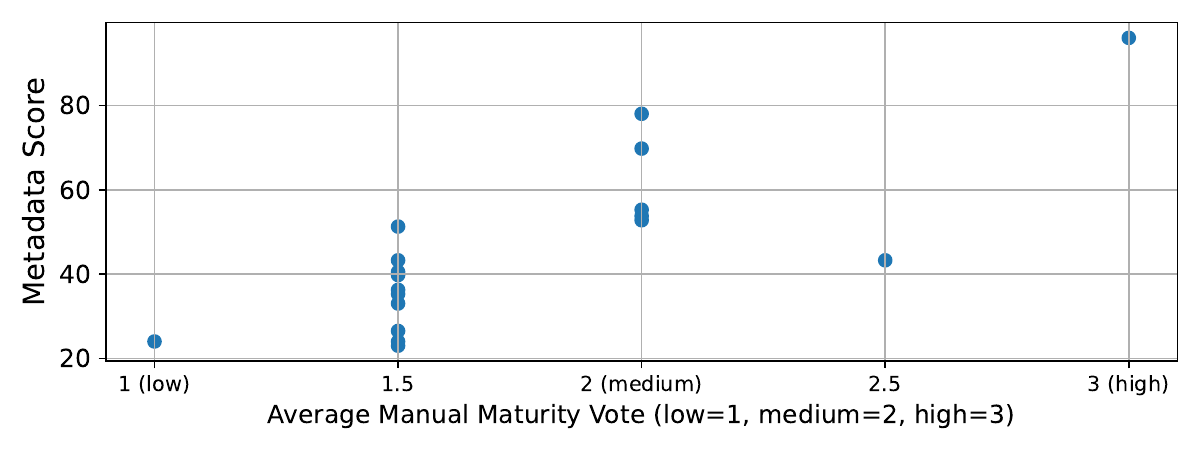}
    \caption{Comparison of maturity scores assigned by manual reviewers and our repository aware scoring approach.}
    \Description{The figure compares maturity scores assigned by manual reviewers with those produced by our repository aware scoring approach.}
    \label{fig:maturity_validation}
\end{figure}

Reviewers also assessed repository maturity. Their judgments were more heterogeneous, but the averaged manual maturity scores show a positive relationship with our metadata score in \Cref{fig:maturity_validation}. This indicates that the metadata component captures repository maturity signals that are also visible to human reviewers, although it should not be interpreted as a definitive maliciousness label.

\section{Discussion}

\paragraph{Skill Marketplaces.}
Our cross-platform analysis shows that agent skill marketplaces largely act as discovery layers on top of GitHub rather than as independent distribution channels. This design increases ecosystem coverage and makes skill publication lightweight, but it also inherits risks from the underlying hosting platform. Broken repository references, renamed accounts, deleted repositories, and private repositories already affected our data collection. More importantly, skills that are referenced rather than mirrored remain vulnerable to repository hijacking when abandoned GitHub namespaces can be re-registered. These findings show that the security of skill marketplaces depends not only on skill contents, but also on how skills are provisioned, authenticated, and archived.

\paragraph{Skill Scanners.}
The scanner results in \roundframe{RQ2} show that malicious classification rates vary substantially across tools and marketplaces. Some scanners classify large fractions of skills as suspicious, yet cross-scanner agreement is low and most flagged skills are detected by only one tool. This indicates that scanner outputs should be interpreted as alerts rather than ground-truth labels. Strictly aggregating these alerts at the repository level further amplifies the problem: repositories containing multiple skills are more likely to be classified as malicious simply because they provide more opportunities for at least one skill to trigger a scanner. This effect can overstate the prevalence of malicious repositories, including for popular projects.

\paragraph{Repository Context.}
The repository-aware analysis in \roundframe{RQ3} addresses this limitation by evaluating scanner-flagged skills within their surrounding project context. Many skills that appear suspicious in isolation are embedded in repositories whose documentation, codebase, and stated purpose align with the skill functionality. As a result, only 15 skill--repository combinations remain suspicious after repository-aware analysis, corresponding to 0.52\% of the evaluated sample. This does not imply that scanners are unnecessary; rather, it shows that users could be provided with more context than isolated skill scanning. Repository context provides an additional signal that can distinguish suspicious behavior from legitimate functionality.

\paragraph{New Ecosystem, Familiar Risks.}
Although agent skills are a new distribution format, many of the observed risks mirror earlier software supply-chain problems. Similar to package managers, container registries, and mobile app stores, skill ecosystems face challenges around authentication, dependency ownership, abandoned projects, and embedded secrets. We found functional credentials in published skills and identified marketplace designs that can enable repository hijacking. These results suggest that the agent skill ecosystem should adopt established supply-chain defenses early, including immutable skill snapshots, stronger publisher authentication, namespace retirement, secret scanning, and clearer provenance metadata.

\paragraph{Practicality of Repository-Aware Scanning.}
Repository-aware analysis is more expensive than scanning a single \texttt{SKILL.md} file, but our cost measurements show that it is practical for periodic marketplace scans. In our sample, repository-context analysis cost around \$0.0097 per skill--repository pair without caching and \$0.0021 with prompt caching. The full sample of 3,000 skills cost approximately \$24 including retries. This makes repository-aware scanning feasible for marketplace operators, especially when combined with caching, incremental scans, and prioritization of newly added or widely installed skills.

\paragraph{Limitations.}
Our repository-aware analysis does not establish a definitive ground truth of maliciousness. It reduces false positives by incorporating repository context, but sophisticated attackers could still craft repositories whose documentation and code appear benign while hiding malicious behavior elsewhere. Our manual validation is limited to a small sample and focuses on visible repository evidence. Moreover, unavailable repositories, failed clones, and dynamic marketplace references limit reproducibility. These limitations reinforce our central conclusion: scanner results should be treated as risk signals that require contextual interpretation, not as definitive labels.

\section{Conclusion}

We presented the largest empirical security analysis of the AI agent skill ecosystem to date, covering 238{,}180 unique skills from three marketplaces and GitHub. Our analysis revealed embedded secrets, marketplace weaknesses, and substantial inconsistencies between existing scanners. Malicious classification rates ranged from 3.8\% to 41.9\%, while agreement was limited: only 0.12\% of commonly analyzed skills were flagged by all five tested scanners.
To address this inconsistency, we proposed a repository aware scanning approach that adds GitHub repository context to skill assessment. Re-evaluating 2{,}887 scanner flagged skill and repository combinations, only 0.52\% remained associated with suspicious repositories. Repository context improved interpretability, as most flagged skills were embedded in repositories whose documentation and code matched the skill functionality, while only 4.2\% showed weak repository linkage.
Finally, we uncovered structural weaknesses in skill marketplaces. We identified repository hijacking risks affecting 121 skills and found that ClawHub exposed sensitive developer metadata, including email addresses associated with GitHub accounts. These findings show that agent skill security cannot be assessed from skill descriptions alone, but requires repository context, stronger provenance guarantees, and established supply chain safeguards.

\section*{Acknowledgments}
The financial support by the Austrian Federal Ministry of Economy, Energy and Tourism, the National Foundation for Research, Technology and Development, and the Christian Doppler Research Association is gratefully acknowledged.

{
  \balance
  \printbibliography

@misc{clawhub_code,
	title        = {{Skill Directory for OpenClaw}},
	author       = {{openclaw}},
	note         = {Accessed: 2026-03-07},
	howpublished = {\url{https://github.com/openclaw/clawhub}}
}

@misc{cisco_skill_scanner,
	title        = {{Skill Scanner: Security Scanner for Agent Skills}},
	author       = {{Cisco AI Defense}},
	year         = 2026,
	note         = {Accessed: 2026-03-07, V2.0.1},
	howpublished = {\url{https://github.com/cisco-ai-defense/skill-scanner}}
}

@misc{claude_code,
  title        = {{GitHub -- Claude Code}},
  author       = {{Anthropic}},
  year         = 2026,
  note         = {Accessed: 2026-03-07},
  howpublished = {\url{https://github.com/anthropics/claude-code}}
}

@misc{openclaw,
	title        = {{OpenClaw -- Personal AI Assistant}},
	author       = {Steinberger, Peter},
	note         = {Accessed: 2026-03-07},
	howpublished = {\url{ https://openclaw.ai/}}
}

@misc{agentskills2025,
	title        = {{Agent Skills: A Simple, Open Format for Giving Agents New Capabilities}},
	author       = {{Agent Skills}},
	year         = 2025,
	note         = {Accessed: 2026-03-07},
	howpublished = {\url{https://agentskills.io/home}}
}

@article{schmotz2026skill,
	title        = {{Skill-Inject: Measuring Agent Vulnerability to Skill File Attacks}},
	author       = {Schmotz, David and Beurer-Kellner, Luca and Abdelnabi, Sahar and Andriushchenko, Maksym},
	year         = 2026,
	journal      = {arXiv preprint arXiv:2602.20156}
}

@inproceedings{schmidt:2025:app_secrets,
	title        = {{Leaky Apps: Large-scale Analysis of Secrets Distributed in Android and iOS Apps}},
	author       = {Schmidt, David and Schrittwieser, Sebastian and Weippl, Edgar},
	year         = 2025,
	booktitle    = {Proc. of ACM CCS},
	doi          = {10.1145/3719027.3765033}
}

@misc{steinberger:2026:clawhub,
	title        = {{ClawHub, the skill dock for sharp agents}},
	author       = {Steinberger, Peter},
	url          = {https://clawhub.ai/},
	note         = {Accessed: 2026-02-26}
}

@misc{virustotal,
  title        = {{VirusTotal}},
  howpublished = {\url{https://www.virustotal.com}},
  note         = {Accessed: 2026-03-09},
  key = {VirusTotal}
}

@misc{kellner:2026:synk,
	title        = {{Snyk Finds Prompt Injection in 36\%, 1467 Malicious Payloads in a ToxicSkills Study of Agent Skills Supply Chain Compromise}},
	author       = {Luca Beurer-Kellner and Aleksei Kudrinskii and Marco Milanta and Kristian Bonde Nielsen and Hemang Sarkar and Liran Tal},
	url          = {https://snyk.io/blog/toxicskills-malicious-ai-agent-skills-clawhub/},
	note         = {Accessed: 2026-02-26},
	date         = {2026-02-05}
}

@misc{lakshmanan:2026:hackernews,
	title        = {{ClawJacked Flaw Lets Malicious Sites Hijack Local OpenClaw AI Agents via WebSocket}},
	author       = {Ravie Lakshmanan},
	url          = {https://thehackernews.com/2026/02/clawjacked-flaw-lets-malicious-sites.html},
	note         = {Accessed: 2026-03-04},
	date         = {2026-02-28}
}

@misc{chandonnet:2026:meta,
	title        = {{Meta AI alignment director shares her OpenClaw email-deletion nightmare:'I had to RUN to my MAC mini'}},
	author       = {Henry Chandonnet},
	url          = {https://www.businessinsider.com/meta-ai-alignment-director-openclaw-email-deletion-2026-2},
	note         = {Accessed: 2026-03-08},
	date         = {2026-02-23}
}

@misc{steinberger:2026:openclaw_github_advisory,
	title        = {{OpenClaw log poisoning (indirect prompt injection) via WebSocket headers}},
	author       = {Peter Steinberger},
	url          = {https://github.com/openclaw/openclaw/security/advisories/GHSA-g27f-9qjv-22pm},
	note         = {Accessed: 2026-03-04},
	date         = {2026-02-14}
}

@misc{cruz:2026:openclaw_keys,
	title        = {{OpenClaw (ex-Moltbot (ex-Clawdbot)): The AI Butler With Its Claws On The Keys To Your Kingdom}},
	author       = {João Cruz},
	url          = {https://www.bitsight.com/blog/openclaw-ai-security-risks-exposed-instances},
	note         = {Accessed: 2026-03-04},
	date         = {2026-02-09}
}

@misc{CVE202559536,
	title        = {{Claude Code' startup trust dialog could lead to Command Execution attack}},
	author       = {CVE-2025-59536},
	url          = {https://www.cve.org/CVERecord?id=CVE-2025-59536},
	note         = {Accessed: 2026-03-08},
	date         = {2025-10-03}
}

@misc{CVE202621852,
	title        = {{Claude Code Leaks Data via Malicious Environment Configuration Before Trust Confirmation}},
	author       = {CVE-2026-21852},
	url          = {https://www.cve.org/CVERecord?id=CVE-2026-21852},
	note         = {Accessed: 2026-03-08},
	date         = {2026-01-21}
}

@misc{shodan:2026:openclaw_instances,
	title        = {{Shodan Search Enging -- OpenClaw}},
	author       = {{Shodan}},
	url          = {https://www.shodan.io/search/report?query=product:openclaw},
	note         = {Accessed: 2026-03-07; Archived at: \url{https://archive.ph/3TNgq}}
}

@misc{fogel:2026:caught_in_the_wild,
	title        = {{Caught in the Wild: Real Attack Traffic Targeting Exposed Clawdbot Gateways}},
	author       = {Ariel Fogel and Eilon Cohen},
	url          = {https://www.pillar.security/blog/caught-in-the-wild-real-attack-traffic-targeting-exposed-clawdbot-gateways},
	note         = {Accessed: 2026-03-07},
	date         = {2026-01-29}
}

@misc{oliveira2026malicious,
	title        = {{Malicious OpenClaw Skills Used to Distribute Atomic macOS Stealer}},
	author       = {Alfredo Oliveira and Buddy Tancia and David Fiser and Philippe Lin and Roel Reyers},
	url          = {https://www.trendmicro.com/en_us/research/26/b/openclaw-skills-used-to-distribute-atomic-macos-stealer.html},
	note         = {Accessed: 2026-03-08},
	date         = {2026-02-23}
}

@misc{skilldirectory,
	title        = {{Agent Skills Directory}},
	author       = {{Skills Directory}},
	url          = {https://www.skillsdirectory.com/},
	note         = {Accessed: 2026-03-04}
}

@misc{skill_specification_mintlify,
	title        = {{Specification -- Agent Skills}},
	author       = {{Mintlify}},
	url          = {https://agentskills.io/specification},
	note         = {Accessed: 2026-03-04}
}

@misc{kellner:2026:snyk_report,
	title        = {{Technical Report: Exploring the Emerging Threats of the Agent Skill Ecosystem}},
	author       = {Luca Beurer-Kellner and Aleksei Kudrinskii and Marco Milanta and Kristian Bonde Nielsen and Hemang Sarkar and Liran Tal},
	url          = {https://github.com/snyk/agent-scan/blob/main/.github/reports/skills-report.pdf},
	date         = {2026-02-05}
}

@misc{skills_sh,
	title        = {{The Agent Skills Directory}},
	author       = {{Vercel Labs}},
	url          = {https://skills.sh},
	note         = {Accessed: 2026-02-26}
}

@misc{schmidt:2026:supplychaininsecurityexposing,
	title        = {{Supply Chain Insecurity: Exposing Vulnerabilities in iOS Dependency Management Systems}},
	author       = {David Schmidt and Sebastian Schrittwieser and Edgar Weippl},
	year         = 2026,
	eprint       = {2601.20638},
	archiveprefix = {arXiv},
}

@inproceedings{gu:2023:software_registries,
	title        = {{Investigating Package Related Security Threats in Software Registries}},
	author       = {Gu, Yacong and Ying, Lingyun and Pu, Yingyuan and Hu, Xiao and Chai, Huajun and Wang, Ruimin and Gao, Xing and Duan, Haixin},
	year         = 2023,
	booktitle    = {Proc of the Symposium on S\&P},
	location     = {San Francisco, CA, USA},
	publisher    = {IEEE},
	doi          = {10.1109/SP46215.2023.10179332},
	isbn         = {978-1-6654-9336-9}
}

@inproceedings{yadmani:2025:s3_secrets,
	title        = {{The File That Contained the Keys Has Been Removed: An Empirical Analysis of Secret Leaks in Cloud Buckets and Responsible Disclosure Outcomes }},
	author       = {El Yadmani, Soufian and Gadyatskaya, Olga and Zhauniarovich, Yury},
	year         = 2025,
	booktitle    = {Proc of the Symposium on S\&P},
	doi          = {10.1109/SP61157.2025.00009}
}

@article{acharya:2025:agenticai,
	title        = {{Agentic AI: Autonomous Intelligence for Complex Goals—A Comprehensive Survey}},
	author       = {Acharya, Deepak Bhaskar and Kuppan, Karthigeyan and Divya, B.},
	year         = 2025,
	journal      = {IEEE Access},
	volume       = 13,
	number       = {},
	pages        = {18912--18936},
	doi          = {10.1109/ACCESS.2025.3532853}
}

@article{pati:2025:agenticai,
	title        = {{Agentic AI: A Comprehensive Survey of Technologies, Applications, and Societal Implications}},
	author       = {Pati, Ashis Kumar},
	year         = 2025,
	journal      = {IEEE Access},
	volume       = 13,
	number       = {},
	pages        = {151824--151837},
	doi          = {10.1109/ACCESS.2025.3585609}
}

@article{deng:2025:artificial,
	title        = {{AI Agents Under Threat: A Survey of Key Security Challenges and Future Pathways}},
	author       = {Deng, Zehang and Guo, Yongjian and Han, Changzhou and Ma, Wanlun and Xiong, Junwu and Wen, Sheng and Xiang, Yang},
	year         = 2025,
	month        = feb,
	journal      = {ACM Comput. Surv.},
	publisher    = {Association for Computing Machinery},
	volume       = 57,
	number       = 7,
	doi          = {10.1145/3716628},
	issn         = {0360-0300},
	url          = {https://doi.org/10.1145/3716628},
	issue_date   = {July 2025},
	articleno    = 182
}

@article{basu:2026:openclaw,
	title        = {{OpenClaw AI chatbots are running amok — these scientists are listening in}},
	author       = {Basu, Mohana},
	year         = 2026,
	journal      = {Nature},
	volume       = 650,
	pages        = 533
}

@article{liu:2026:agent,
	title        = {{Agent Skills in the Wild: An Empirical Study of Security Vulnerabilities at Scale}},
	author       = {Liu, Yi and Wang, Weizhe and Feng, Ruitao and Zhang, Yao and Xu, Guangquan and Deng, Gelei and Li, Yuekang and Zhang, Leo},
	year         = 2026,
	journal      = {arXiv preprint arXiv:2601.10338}
}

@article{ling:2026:agent,
	title        = {{Agent Skills: A Data-Driven Analysis of Claude Skills for Extending Large Language Model Functionality}},
	author       = {Ling, George and Zhong, Shanshan and Huang, Richard},
	year         = 2026,
	journal      = {arXiv preprint arXiv:2602.08004}
}

@article{bhardwaj:2026:formal,
	title        = {{Formal Analysis and Supply Chain Security for Agentic AI Skills}},
	author       = {Bhardwaj, Varun Pratap},
	year         = 2026,
	journal      = {arXiv preprint arXiv:2603.00195}
}

@inproceedings{bugiel:2011:amazonia,
	title        = {{AmazonIA: when elasticity snaps back}},
	author       = {Bugiel, Sven and N\"{u}rnberger, Stefan and P\"{o}ppelmann, Thomas and Sadeghi, Ahmad-Reza and Schneider, Thomas},
	year         = 2011,
	booktitle    = {Proc. of ACM CCS},
	doi          = {10.1145/2046707.2046753},
	isbn         = 9781450309486,
}

@inproceedings{shu:2017:study,
	title        = {{A Study of Security Vulnerabilities on Docker Hub}},
	author       = {Shu, Rui and Gu, Xiaohui and Enck, William},
	year         = 2017,
	booktitle    = {Proc. of the ACM on Conference on Data and Application Security and Privacy},
	location     = {Scottsdale, Arizona, USA},
	doi          = {10.1145/3029806.3029832},
	isbn         = 9781450345231,
}

@article{zhang:2025:killing,
	title        = {{Killing Two Birds with One Stone: Malicious Package Detection in NPM and PyPI using a Single Model of Malicious Behavior Sequence}},
	author       = {Zhang, Junan and Huang, Kaifeng and Huang, Yiheng and Chen, Bihuan and Wang, Ruisi and Wang, Chong and Peng, Xin},
	year         = 2025,
	month        = apr,
	journal      = {ACM Transactions on Software Engineering and Methodology},
	publisher    = {Association for Computing Machinery},
	volume       = 34,
	number       = 4,
	doi          = {10.1145/3705304},
	issn         = {1049-331X},
	url          = {https://doi.org/10.1145/3705304},
	issue_date   = {May 2025},
	articleno    = 104
}

@misc{artifact:skill_scripts,
	title        = {{GitHub -- Skill Scripts Table}},
  author={Holzbauer, Florian and Schmidt, David and Gegenhuber, Gabriel and Schrittwieser, Sebastian and Ullrich, Johanna},
	year         = 2026,
    url          = {https://github.com/holzsec/repository-context-agentskills/blob/main/rq1_skill_dataset/tables/skill_content.pdf}
}

}

\appendix

\end{document}